# Diffraction Interference Induced Superfocusing in Nonlinear Talbot Effect

Dongmei Liu,[1] Yong Zhang,[1,*] Jianming Wen,[1,2,†] Zhenhua Chen,[1] Dunzhao Wei,[1] Xiaopeng Hu,[1] Gang Zhao,[1] S. N. Zhu,[1] and Min Xiao[1,3,‡]

[1]National Laboratory of Solid State Microstructures, College of Engineering and Applied Sciences, School of Physics, Nanjing University, Nanjing 210093, China

[2]Department of Applied Physics, Yale University, New Haven, Connecticut 06511, USA

[3]Department of Physics, University of Arkansas, Fayetteville, Arkansas 72701, USA

Abstract:
We report a simple, novel subdiffraction method, i.e. diffraction interference induced superfocusing in second-harmonic (SH) Talbot effect, to achieve focusing size of less than $\lambda_{pump}/8$ without involving evanescent waves or subwavelength apertures. By tailoring point spread functions with Fresnel diffraction interference, we observe periodic SH subdiffracted spots over a hundred of micrometers away from the sample. Our demonstration is the first experimental realization of the Toraldo Di Francia's proposal pioneered 60 years ago for superresolution imaging.

Focusing of a light beam into an extremely small spot with a high energy density plays an important role in key technologies for miniaturized structures, such as lithography, optical data storage, laser material nanoprocessing and nanophotonics in confocal microscopy and superresolution imaging. Because of the wave nature of light, however, Abbe [1] discovered at the end of 19th century that diffraction prohibits the visualization of features smaller than half of the wavelength of light (also known as the Rayleigh diffraction limit) with optical instruments. Since then, many efforts have been made to improve the resolving power of optical imaging systems, and the research on overcoming the Abbe-Rayleigh diffraction limit has become an energetic topic (for recent reviews see Refs. [2-4]).

Historically, an early attempt to combat the diffraction limit can be traced back to the work by Ossen [5] in 1922, in which he proved that a substantial fraction of emitted electromagnetic energy can be squeezed into an arbitrarily small solid angle. Inspired by the concept of super-directivity [6], in his seminal 1952 paper Toraldo Di Francia [7] suggested that a pupil design provides an accurately tailored subdiffracted spot by using a series of concentric apertures with different phases. Based on the mathematical prediction that band-limited functions are capable of oscillating faster than the highest Fourier components carried by them (a phenomenon now known as superoscillation) [8], Berry and Popescu [9] in their recent theoretical analysis pointed out that subwavelength localizations of light could be obtained in Talbot self-imaging [10,11] under certain conditions. With the use of a nanohole array, Zheludev's group demonstrated the possibility to focus light below the diffraction limit [12,13]. By using a sequence of metal concentric rings with subwavelength separations, they further reported well-defined, sparsely distributed subdiffracted light localizations in a recent optical superoscillating experiment [14]. Despite the newly developed quantum imaging [15] and quantum lithography [16] techniques also allow the formation of sub-Rayleigh diffracted spots, the severe reliance on specific quantum entangled states and sophisticated measurement devices limits their practical applications.

By exploring evanescent components containing fine details of an electromagnetic field distribution, researchers working in near-field optics have invented powerful schemes, such as total internal reflectance microscopy [17] and metamaterial-based superlens [18,19], to overcome the barrier of the diffraction limit. Most near-field techniques operate at a distance extremely close (typically hundreds of nanometers) to the object in order to obtain substantial subdiffracted spots. Since these techniques cannot image an object beyond one wavelength, they are not applicable to image into objects thicker than one wavelength, which greatly limits their applicability in many situations. There also exists a broad category of *functional* super-resolution imaging techniques which use clever experimental tools and known limitations on the matter being imaged to reconstruct the super-resolution images. The representative ones include stimulated emission depletion [20], spatially-structured illumination microscopy [21], stochastic optical reconstruction microscopy [22], and super-resolution optical fluctuation imaging [23].

Here we introduce yet another alternative scheme, i.e. diffraction interference induced superfocusing in nonlinear Talbot effect [24,25], to achieve subdiffraction by exploiting the phases of the second-harmonic (SH) fields generated from a periodically-poled LiTaO$_3$ (PPLT) crystal. The poling inversions in the PPLT crystal, typically with a period of few micrometers, make the SH waves generated in the negative domains possess a π phase shift relative to those in the positive domains. The destructive interference between these two generated SH waves in the Fresnel diffraction region shrinks the point spread functions below the diffraction limit and leads to subwavelength focused spots, closely resembling the idea as suggested by Toraldo Di Francia sixty years ago [7]. Besides, because of the phase matching, the generated SH signals are automatically band-limited, which is a key ingredient for realizing superoscillations [9]. These two unique and coexisting features distinguish the current scheme from all previous works that involve either evanescent waves, metal nanostructures, luminescent objects or quantum states. Our demonstration can be considered as the first experimental realization of the Toraldo Di Francia's proposal for subdiffraction [7] and superresolution with superoscillations [9]. This method allows to produce subdiffracted SH spots over 100 μm easily, and has, in principle, no fundamental lower bound to limit the focusing ability. As such, we have observed superfocused SH spots with the size of less than one quarter of the SH wavelength ($\lambda_{SH}/4$) at the distance of tens of micrometers away, which is comparable to the superoscillating experiment [14], but without employing subwavelength metal nanoholes. We thus expect our imaging technique to provide a super-resolution alternative for various applications in photolithography, medical imaging, molecular imaging, as well as bioimaging.

In our proof-of-principle experiment, the periodic domain structures of the PPLT crystal help create subwavelength foci with prescribed sizes and shapes, as the SH waves with different phases propagate freely and interfere destructively. The achievable subdiffraction patterns depend on parameters such as the periodicity of domain structures, sizes of the domain structures, and the propagation distance. Experimentally, superfocused SH spots with sizes of less than $\lambda_{pump}/8$ have been recorded at 27.5 μm away from the sample. In comparison with the superoscillatary experiment [14] using a binary-amplitude metal mask, the current scheme explores the $\pi$ phase difference between the SH fields generated from inside and outside of the domains, respectively. Note that such π phase shift does not exist in grating-based Talbot effect and linear superfocusing systems. Besides, the structure of the PPLT crystal does not involve complicated nano-fabrications and has large-scale parameters than light wavelengths.

Similar to our previous studies on SH Talbot effects [24, 25], the superfocusing setup is schematically shown in Fig. 1. A femtosecond mode-locked Ti:sapphire laser was operated at a wavelength of $\lambda_{pump}$ = 900 nm as the fundamental input field. The pulse width was about 75 fs with a repetition rate of 80 MHz. As illustrated in Fig. 1, the fundamental pump laser was first shaped by a pinhole and focusing lens to produce a near-parallel beam with a spot size of ~100 μm, and then directed into a 2D squarely-poled LiTaO$_3$ slice along the z axis with its polarization parallel to the x axis of the crystal. The sample with the size of 20 mm (x) ×20 mm (y) ×0.5 mm (z) was placed on the microscope stage, and the SEM image of its domain structures (with the period of a = 5.5 μm and the duty cycle of ~35%) is depicted in the inset of Fig. 1. Despite the LiTaO$_3$ crystal has a space group of 3m (C$_{3v}$), only the $d_{21}$ component contributes to the SH generation in the current experimental configuration. After the sample, an objective lens (×100) with a high numerical aperture of NA = 0.7 was used to magnify the generated SH intensity patterns ($\lambda_{SH}$ = 450 nm). To remove the near-infrared fundamental field, a bandpass filter was placed between the objective lens and the CCD camera. The SH patterns at different imaging planes were recorded by moving the microscope stage along the SH propagation direction, which was controlled by a precision translation stage.

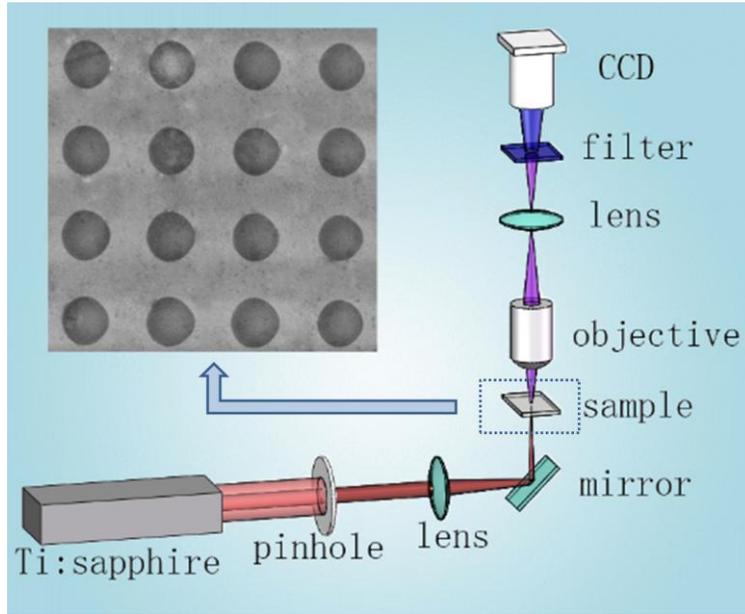

FIG. 1 Schematic diagram of the experimental setup. The sample is placed at the focal plane of the lens. The SH patterns at different imaging planes are recorded by a CCD camera through moving the microscope stage. Inset is the SEM image of the 2D squarely-poled LiTaO$_3$ slice with a period of 5.5 μm and duty cycle of ~35%.

Characteristic SH field patterns recorded at different distances (Z) away from the output surface of the sample are presented in Fig. 2, representing a variety of "photonic carpets" in the Fresnel diffraction region. As indicated in Figs. 2(a)-2(f), the diffraction patterns change dramatically along with the focus being moved away from the crystal. The primary Talbot self-imaging was observed at Z = 132.3 μm [Fig. 2(f)], which is well consistent with the theoretically calculated SH Talbot length [24,25] of $Z_t = 4a^2/\lambda_{pump}$ = 134.4 μm. One previously unconfirmed feature appears at about 1/2 Talbot length where a square array-like SH self-image is laterally shifted by half the width of the domain period [comparing Fig. 2(c) with Fig. 2(f)]. In fractional Talbot planes, one can see complicated and beautiful patterns, which result from the diffraction interferences of the SH waves. In proximity to the end face of the sample [Figs. 2(a)-2(d)], the SH waves form periodic focusing spots at the center of each unit. The focusing size varies with the propagation distance Z and superfocusing occurs at certain planes. At some other planes [e.g., 4/5 fractional Talbot plane as shown in Fig. 2(e)], the subdiffraction focusing spots disappear due to destructive interference. By carefully examining the patterns, one can find that the detailed structures in every single unit are very sensitive to the observation distance, especially when close to the sample. For example, as the observation plane moves from Z = 3.5 μm [Fig. 2(a)] to Z = 4.6 μm [Fig. 2(b)], the rings at the center shrink and the fractal array at the corner evolves. The key factor is that the phases of the SH waves develop sensitively along the propagation distance.

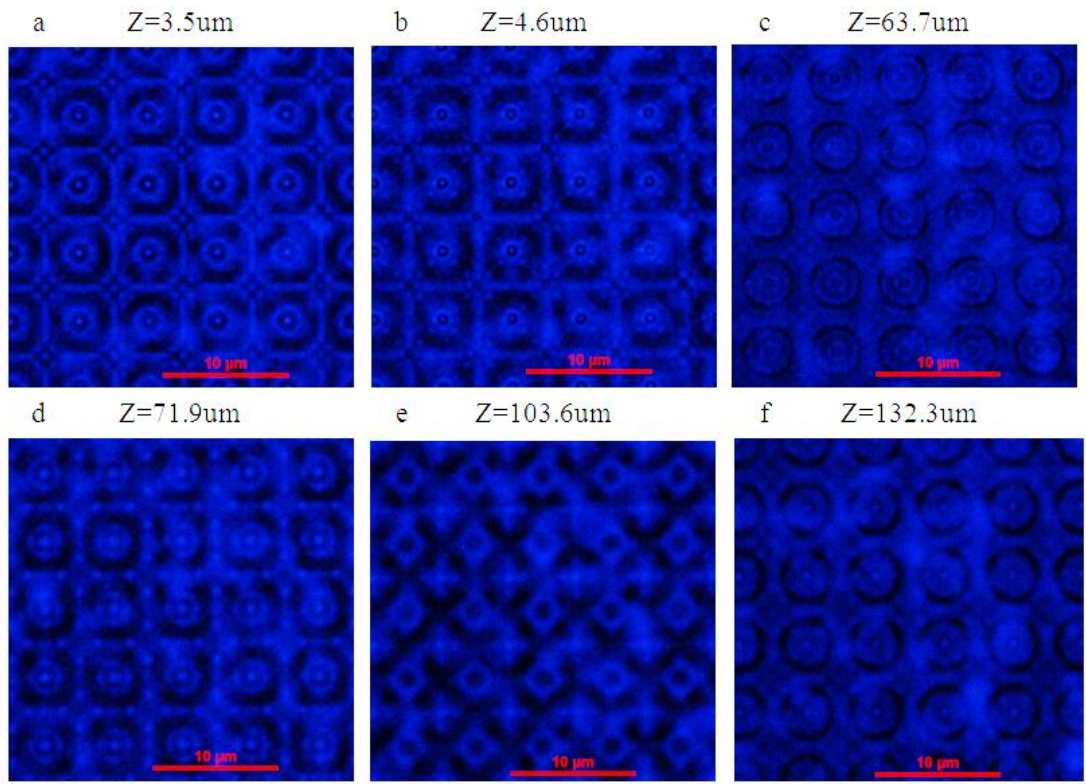

FIG. 2 Recorded images of the second-harmonic patterns with a conventional optical microscope at different Talbot planes.

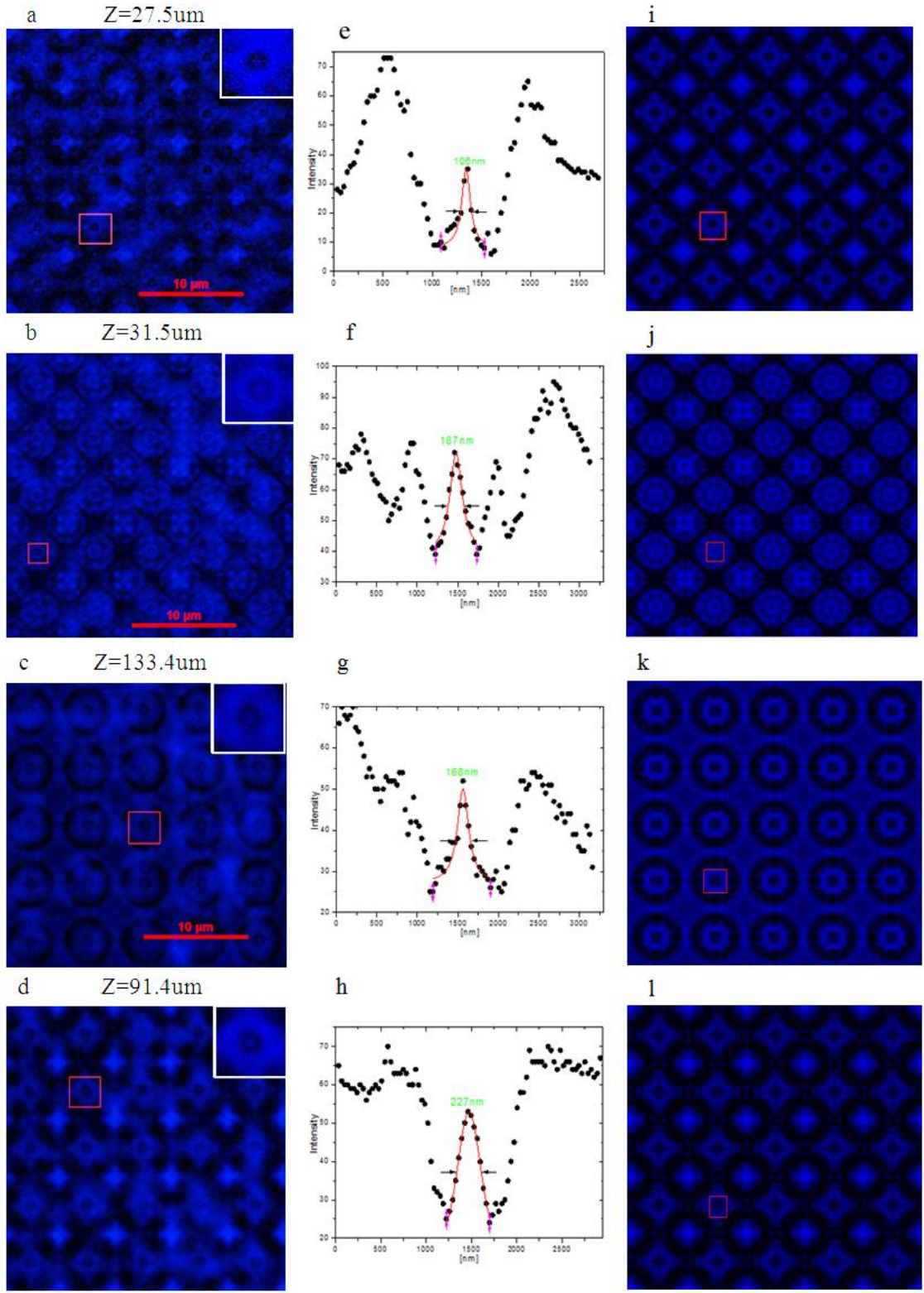

FIG. 3 Typical measured results of superfocusing. a-d are experimentally recorded SH patterns at different observation distances, where the insets are the enlarged images of the selected focusing spots. The cross-sections of the selected focusing spots in a-d are given, respectively, in e-h, whose centers are fitted with a Lorentzian (e-g) or Gaussian (h) lineshape. The black dots are the measured SH intensities. i-l are theoretical simulations corresponding to a-d, respectively.

The superfocusing feature was carefully measured and analyzed in our experiment. Figure 3 shows some typical images with subwavelength focused spots. For $\lambda_{SH}$ = 450 nm, at a distance of Z = 27.5 μm a subdiffracted spot was identified with a full-width-at-the-half-maximum (FWHM) of 106 nm [see the enlarged area in Fig. 3(a)], i.e. 0.117$\lambda_{pump}$ or 0.235$\lambda_{SH}$, which surpasses the result reported for the super-oscillatory lens (0.29$\lambda_{pump}$; a focal spot of 185 nm in diameter for a wavelength of 640 nm, Ref. [14]). The recorded cross-section of the spot is shown in Fig. 3(e) without any data post-processing, which is well fitted with a Lorentzian lineshape. The background is resulted from the imperfect domain structures. Due to the imperfections, such as defects in the crystal and the nonuniform domain structures, not all the theoretically predicted superfocusing spots were observable in the experiment. At Z = 31.5 μm the selected subdiffraction spot [Fig. 3(b)] has a FWHM of 187 nm [Fig. 3(f)]. We notice that in the current scheme, the cross-section profile of the measured subdiffracted spot fits better with a Lorentzian line shape rather than with a Gaussian shape, which is in contrast to those focusing spots obeying the diffraction limit. For instance, at the distance of Z = 91.4 μm, a focusing spot with a size of 227 nm (close to the diffraction limit) can be well fitted by a Gaussian curve [Figs. 3(d) and 3(h)]. In the present experiment, the largest distance where we can still find superfocusing spots is Z = 133.4 μm [almost at the primary SH Talbot plane, Fig. 3(c)] and these subwavelength spots with a FWHM down to 168 nm [Fig. 3(g)] well follow a Lorentzian profile.

To theoretically confirm that the superfocusing feature in our experiment is indeed formed by the destructive interference of propagating SH waves, we performed numerical simulations using the angular spectrum method. We used the same parameters of the sample to model the "aperture function", and also took into account the π phase shift of the SH waves generated in the negative domains. After a propagation distance z, the diffracted SH field is computed by the Rayleigh–Sommerfeld diffraction formula as [25]:

$$U(x,y,z) = \int_{-f_m}^{f_m}\int_{-f_m}^{f_m} A_0(f_x,f_y)\exp[ikz\sqrt{1-(\lambda f_x)^2-(\lambda f_y)^2}]\exp[i2\pi(f_x x + f_y y)]df_x df_y,$$

where $A_0(f_x,f_y)$ is the angular-spectrum representation of the sample aperture function at z = 0. In the model, the SH field is simplified to be a plane wave, and the integration limits are bounded by the phase-matching condition in the range of $[-f_m, f_m]$. For the self-images shown in Figs. 3(a)-3(d), the simulations predict the same patterns as depicted in Figs. 3(i)-3(l), respectively. The simulations yield the focused spot sizes of 118 nm, 191 nm, 185 nm, and 225 nm at Z = 27.5 μm, 31.5 μm, 133.4 μm, and 91.4 μm, respectively, which indicate excellent agreements with the corresponding experimental data. Our simulations also reveal that the width of the domain walls is a very important parameter for the imaging pattern. We find, both theoretically and experimentally, that wide domain walls (>1 μm) may completely change the image patterns and eliminate the superfocusing phenomenon. With the high-quality samples used in the experiment, the computed patterns by using the above model without considering the domain wall width can well match the experimental results. This implies that the domain walls in those samples are narrow enough to be negligible in the process of image formations.

To further verify the observed subdiffraction effect, we have chosen another hexagonally-poled PPLT structure with a period of 9 μm and the duty cycle ~30% (previously used for the illustration of SH Talbot effect [24,25]). The input pump laser was still at the wavelength of 900 nm. As an example, the SH self-image [Fig. 4(a)] at a distance of 3 μm away from the PPLT crystal was recorded and analyzed. As expected, a superfocused spot size of 168 nm was identified, as shown in Fig. 4(b).

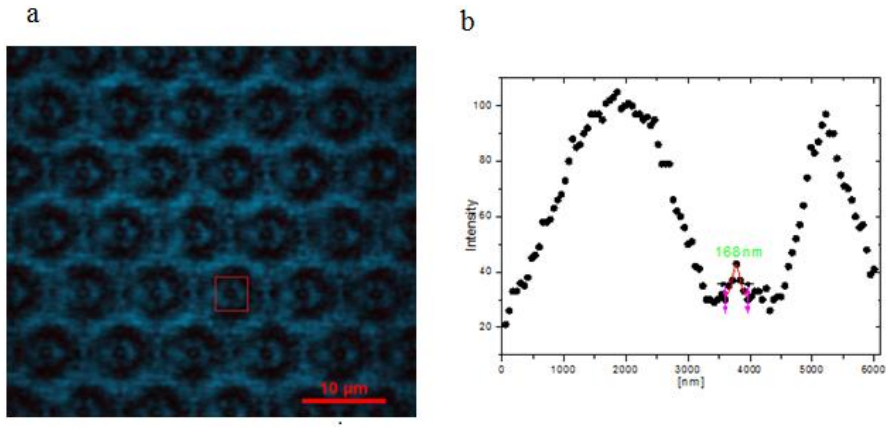

FIG. 4. Superfocusing with a hexagonally-poled PPLT crystal. a is the SH self-image recorded at a distance of z = 3 μm. The SH wavelength is 450 nm. b is the cross-sectional profile of the superfocusing spot marked in red square in a. Its curve is well fitted with a Lorentzian model with the FWHM of 168 nm.

Subwavelength focusing with such a square array or hexagonal array PPLT could allow light to be squeezed into a spatial region with the scale of less than a quarter of the SH wavelength (i.e., less than one eighth of the fundamental wavelength!), thereby opening new avenues for studying light-matter interactions, single-molecule sensing, nanolithography, and nanoscale imaging. By optimizing the sample parameters (such as the periodicity, domain structures, and propagation distance), it is possible to further shrink the focused spot size down to tens of nanometers, which would then become comparable with those functional super-resolution imaging techniques [20-23]. Moreover, due to the excellent electro-optic characteristics of the PPLT crystals, one may continuously tune the phases of the SH waves produced in the crystal structure, and control the interference and focusing of the generated SH waves in the far field by applying an electric field [26]. Subdiffraction imaging holds many exciting promises in many areas of science and technologies. Extensions of the current method to optical microscope may improve the resolving power down to nanometer scale, which would become very useful for non-invasive subwavelength biomedical imaging. Another potential application is in optical lithography at ultra-small scales, which is the key for scaling down integrated circuits in high-performance optoelectronics. Optical data storage and biosensing may also benefit from this promising scheme to process information within an ultra-small volume, and thereby increase storage density or sensing resolution.

In summary, we have proposed and demonstrated a simple way to reduce the point spread function below $\lambda_{pump}/8$ in SH Talbot effect with the periodically-poled LiTaO$_3$ crystal. The method involves neither evanescent waves nor subwavelength structures in the object. Through the destructive interference, the subdiffracted SH spots can be observed up to 133.4 μm away from the sample for $\lambda_{pump}$ = 900 nm. The numerical simulations have confirmed the experimental results with excellent agreements. Our work can be considered as the first realization of the proposals made by Toraldo di Francia [7] and Berry & Popescu [9]. Furthermore, our investigation can potentially have a wide range of applications including subwavelength imaging, as a mask for biological molecule imaging, optical lithography and focus devices.

This work was supported by the National Basic Research Program of China (Nos. 2012CB921804 and 2011CBA00205), the National Science Foundation of China (Nos. 11274162, 61222503, 11274165 and 11021403), the New Century Excellent Talents in University, and the Priority Academic Program Development of Jiangsu Higher Education Institutions (PAPD).